# Discovery of delta Scuti variables in eclipsing binary systems II. Southern TESS field search

F. Kahraman Aliçavuş [ORCID],[1,2]⋆ Ç. G. Çoban,[3] E. Çelik,[3] D. S. Dogan,[4] O. Ekinci[3] and F. Aliçavuş[1,2]

[1]*Çanakkale Onsekiz Mart University, Faculty of Sciences, Physics Department, 17100 Çanakkale, Türkiye*  
[2]*Çanakkale Onsekiz Mart University, Astrophysics Research Center and Ulupınar Observatory, TR-17100 Çanakkale, Türkiye*  
[3]*Çanakkale Onsekiz Mart University, School of Graduate Studies, Department of Space Sciences and Technologies, TR-17100 Çanakkale, Türkiye*  
[4]*Çanakkale Onsekiz Mart University, Faculty of Sciences, Department of Space Sciences and Technologies, TR-17100 Çanakkale, Türkiye*

## ABSTRACT

The presence of pulsating stars in eclipsing binary systems (EBs) makes these objects significant since they allow us to investigate the stellar interior structure and evolution. Different types of pulsating stars could be found in EBs, such as $\delta$ Scuti variables. $\delta$ Scuti stars in EBs have been known for decades, and the increasing number of such systems is important for understanding pulsational structure. Hence, in this study, research was carried out on the southern TESS field to discover new $\delta$ Scuti stars in EBs. We produced an algorithm to search for detached and semidetached EBs considering three steps: the orbital period ($P_{\rm orb}$)'s harmonics in the Fourier spectrum, skewness of the light curves, and classification of UPSILON program. If two of these steps classify a system as an EB, the algorithm also identifies it as an EB. The TESS pixel files of targets were also analysed to see whether the fluxes were contaminated by other systems. No contamination was found. We researched the existence of pulsation through EBs with a visual inspection. To confirm $\delta$ Scuti-type oscillations, the binary variation was removed from the light curve, and residuals were analysed. Consequently, we identified 42 $\delta$ Scuti candidates in EBs. The $P_{\rm orb}$, $L$, and $M_{\rm V}$ of systems were calculated. Their positions on the H–R diagram and the known orbital-pulsation period relationship were analysed. We also examined our targets to see if any of them showed frequency modulation with the orbital period and discovered one candidate of tidally tilted pulsators.

## 1 INTRODUCTION

There are many different kinds of variable stars that significantly contribute to the examination of stellar structure and evolution. The pulsating stars are one of these variables. The Hertzsprung–Russell (H–R) diagram is covered by various kinds of pulsating variables having distinct evolutionary statutes. The oscillation frequencies of pulsating stars could be used to probe the stellar interior structure with a method known as Asteroseismology. Asteroseismology can reveal some significant properties of stars, such as core overshooting and internal rotation (e.g. Christophe et al. 2018; Townsend, Goldstein & Zweibel 2018; Walczak et al. 2019).

One of the notable pulsating star groups is the $\delta$ Scuti variable stars. The $\delta$ Scuti variables have been known for decades, with their short oscillation period ranging approximately from 18 mins to 8 h (Aerts, Christensen-Dalsgaard & Kurtz 2010). The $\delta$ Scuti stars are A-F type variables, with the luminosity type changing between dwarf and giant (Chang et al. 2013). These variables exhibit low degree pressure, gravity, and mixed oscillation modes with an amplitude value generally lower than $0^m.1$ in *V*-band (Chang et al. 2013). As the $\delta$ Scuti variables show oscillation in different evolutionary statuses and are placed in the region where the radiative core turns into convective core (Aerts et al. 2010), these variables are quite crucial systems to examine the progress of stellar evolution and structure.

The other significant variables are the binary systems. Particularly, the eclipsing binary stars are unique objects that provide the most accurate fundamental stellar parameters such as radius (*R*) and mass (*M*). The accuracy in these parameters could be as low as 1 per cent (Torres, Andersen & Giménez 2010; Southworth 2013). Precisely determined fundamental stellar parameters are essential for analysis to get sensitive results. It is also known that binary stars may consist of pulsating stars such as $\delta$ Scuti pulsators (Breger 2000; Kahraman Aliçavuş et al. 2017; Liakos & Niarchos 2017). The eclipsing binary systems with pulsating component(s) are powerful tools that provide both precise fundamental stellar parameters and also a way to probe the stellar interior via asteroseismology. That makes the eclipsing binaries with a pulsating component extremely important.

Many studies have been carried out on eclipsing binary systems with pulsating components and revealed their properties (i.e. Handler et al. 2020; Rappaport et al. 2021; Kahraman Aliçavuş et al. 2023). In some cases, the catalogue of $\delta$ Scuti stars in binaries were given (Kahraman Aliçavuş et al. 2017; Liakos & Niarchos 2017). In Kahraman Aliçavuş et al. (2017), it was shown that there were around 90 eclipsing binaries with $\delta$ Scuti components and this number is increasing with the discovery of new candidates with the help of space telescopes such as Transiting Exoplanet Survey Satellite (TESS; Ricker et al. 2014; Kahraman Aliçavuş et al. 2022). It is known that binarity has a significant influence on pulsations (Kahraman Aliçavuş et al. 2017; Liakos & Niarchos 2017). Recently discovered star group, tidally tilted pulsators, are one of the samples. In these systems, the pulsation axis is changed and aligned with the orbital axis due to tidal forces (Handler et al. 2020; Rappaport et al. 2021). Discoveries reveal other properties of the $\delta$ Scuti stars in eclipsing binaries, and to deeply understand the processes occurring in the $\delta$ Scuti stars detailed studies of these kinds of objects are necessary. Therefore, in this study, we present our research on the southern TESS field to find new $\delta$ Scuti star candidates in eclipsing binaries. In our first paper, we carried out a similar research on the Northern hemisphere but with a visual examination (Kahraman Aliçavuş et al. 2022). However, in this study, we prepared an algorithm that takes into account the structure of frequencies and skewness of data and also uses the UPSILON program, which is an automatic classification program for variable stars (Kim & Bailer-Jones 2016).

⋆ E-mail: filizkahraman01@gmail.com

The paper is organized as follows. Information about the used photometric data is given in Section 2. The identification of eclipsing variability is presented in Section 3. Determination of orbital period and pulsational analysis are introduced in Section 4. In Section 5, we give the calculation of some basic parameters. Discussion and conclusions are given in Set. 6.

## 2 PHOTOMETRIC DATA

In this study, we aim to identify new $\delta$ Scuti variables in eclipsing binary systems. For this purpose, as a continuation of our previous work (Kahraman Aliçavuş et al. 2022), we carried out research on the southern hemisphere data of TESS. TESS is a space telescope that was launched in 2018 for discovery of new exoplanets in close, bright stars (Ricker et al. 2014). TESS observes the sky by dividing it into sectors that consist of around 27 d of observation. TESS provides two types of data known as long and short cadences. The short cadence (SC) data have been taken in 120-s cadence, while the long cadence (LC) data have been taken in 600- and 1800-s cadences. These TESS data are presented in the Barbara A. Mikulski Archive for Telescopes (MAST).[1]

We collected the SC data of the southern objects because the Nyquist frequency of SC data reaches up to $360\,d^{-1}$. As $\delta$ Scuti variables exhibit the oscillation frequencies in general between $\sim 5$ and $80\,d^{-1}$ (Chang et al. 2013), SC data is more suitable for the discoveries of $\delta$ Scuti stars in eclipsing binaries. Hence, we took into account the TESS SC data in this study. TESS provides two types of fluxes, simple aperture photometry (SAP), and pre-search data conditioning SAP (PDCSAP). In our study, the SAP data were used since it has lower uncertainty and PDCSAP fluxes of eclipsing binary (EB) systems mostly show asymmetry.

## 3 IDENTIFICATION OF ECLIPSING VARIABILITY

We investigated $\delta$ Scuti stars through EB systems. Therefore, we needed to identify EB systems in the Southern hemisphere. We developed an algorithm that detects EB systems using the TESS light curve. Since the $\delta$ Scuti stars have been found in detached and semidetached eclipsing binaries (Kahraman Aliçavuş et al. 2017; Liakos & Niarchos 2017), we only aimed to identify detached and semidetached eclipsing binaries with our algorithm.

---
[1]https://mast.stsci.edu

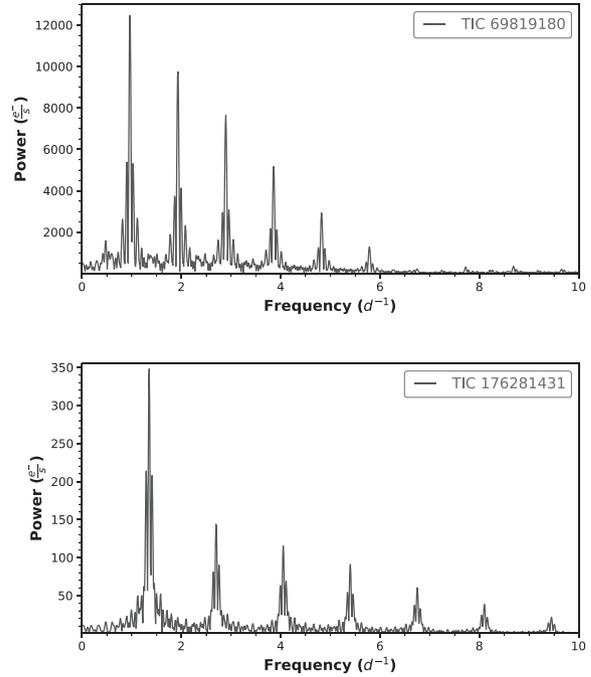

**Figure 1.** Top panel: Frequency spectrum of an EB system (KX Aqr, TIC 69819180, Kazarovets et al. Kazarovets et al., 1999). Bottom panel: Frequency spectrum of a RR Lyrae system (GK Cet, TIC 176281431, Kinemuchi et al. Kinemuchi et al., 2006).

The algorithm was developed using the PYTHON programming language and utilizes the LIGHTKURVE (Lightkurve Collaboration 2018) and ASTROPY (Astropy Collaboration 2018) libraries to download 120-second SAPFLUX data from the Mikulski Archive for Space Telescopes (MAST) data base. Some targets have a few TESS sectors' data, for these, we only took into account one/two sector data for the analysis. Those TESS data are then converted into magnitudes using the Pogson formula.

The effective temperature ($T_{eff}$) and surface gravity ($\log g$) of $\delta$ Scuti stars range between 6300–8500 K, and 3.2–4.3, respectively (Rodríguez & Breger 2001). We examined EB systems for the presence of $\delta$ Scuti stars considering the general $T_{eff}$ range of $\delta$ Scuti variables with a possible error on $T_{eff}$. In the study, the $T_{eff}$ values of the systems were automatically taken from the TESS Input Catalogue (TIC; Stassun et al. 2019), and we searched for the systems that have $T_{eff}$ between $\sim 5300$ and $13\,000$ K considering a 15–30 per cent error in TIC $T_{eff}$. This is because the TESS satellite is designed to observe much cooler systems, and the systems in interest are binaries; hence, there would be errors in the measured $T_{eff}$ values.

After collecting the data of southern stars from MAST and considering the stars that have the $T_{eff}$ values in the range of $\sim 5300$–$13\,000$ K, we created an algorithm to identify EBs. The algorithm uses three different methods to identify EB systems. The first method is the identification of EB systems using the harmonic frequencies of orbital periods. The algorithm obtains the frequency spectra of the systems by applying Fourier analysis to the TESS data. The frequency spectra of EB stars mostly exhibit the harmonic frequencies of the orbital period, which have the highest amplitude, as shown in the top panel of Fig. 1 for one EB sample from our list. Therefore, using the developed algorithm, six consecutive frequencies with the highest amplitude were detected on the frequency spectrum of individual stars. The differences between the six consecutive frequencies were controlled to identify the harmonic frequencies, taking into account a 1 per cent tolerance. If an eclipsing binary system has an orbital period approximately lower than 10 d, that demonstrates frequencies similar to those shown in Fig. 1. However, there are some EBs that have longer orbital periods and show only a few eclipses (or one) in one TESS sector. In this case, the frequency(s) of the orbital period mostly cluster below 1 d$^{-1}$. Therefore, if there are at least four harmonic frequencies and/or frequencies below 1 d$^{-1}$ plus harmonic frequency(s), the algorithm classifies the system as an EB.

Although the first step of the algorithm works well, we noticed that it has difficulty distinguishing RR Lyrae stars from the EBs. As can be seen from Fig. 1, the frequency distribution of EB and RR Lyrae systems shows similarities. If we examine the figures, we can see that both systems exhibit harmonic frequencies. Due to this, RR Lyrae stars cannot be distinguished very accurately from EB systems using harmonic frequencies. Therefore, as another method in our algorithm, the skewness of the light curve was analysed. The skewness value is almost zero in light curves with a symmetric distribution. However, in light curves that do not have a symmetric distribution, such as in most RR Lyrae stars, the skewness value takes positive and negative values. The algorithm also determines whether the system is an EB or non-EB based on the skewness value calculated from the light curve data of the systems.

Another method we included in our algorithm is a machine learning model. Machine learning algorithms are trained on data sets and then used to identify new data in those sets. Therefore, we use a machine learning model called UPSILON (Kim & Bailer-Jones 2016) to determine the type of systems in our study. This model has been trained on different types of stars such as RR Lyrae, δ Scuti, and EBs. The model's success rate in detecting detached and semidetached binary systems is 91 and 74 per cent, respectively (Kim & Bailer-Jones 2016). The model determines the type of system using the time, brightness, and magnitude errors.

The algorithm identifies the system as EB or non-EB using the harmonic frequencies, UPSILON, and skewness methods. If at least two methods identify the system as an EB that system is classified to be an EB. Our developed algorithm was tested on a dataset consisting of 50 stars of RR Lyrae, δ Scuti, EB (detached and semidetached), γ Doradus, and rotational variables. As we focused on the identification of the EB systems, the success of our algorithm in the determination of EBs was very important. Hence, around half of the sample is EBs with an orbital period in the range of 0.64–20.18 d. The list of the test sample is given in Table A1. In our examination of samples, the algorithm detected EB-type stars with 100 per cent accuracy. When examining all methods used in the algorithm separately, the skewness and UPSILON methods identified EBs 100 per cent correct, while the success of the harmonic frequency method is 72 per cent.

After testing our algorithm, we used it to identify EB systems in the southern TESS field. The algorithm analysed approximately 18 000 southern stars from the TESS satellite[2] and classified a group of stars as EB. Since we aim to find δ Scuti variables in EBs, after the determination of EBs, we visually inspected the maximums of light curves to find the oscillation-like variations. As a result, 150 EB systems were determined as candidates of pulsators in EBs. To be sure about the δ Scuti-type variation in these EB systems, we carried out a pulsational analysis.

## 4 DETERMINATION OF THE ORBITAL PERIODS AND PULSATIONAL ANALYIS

In our previous analysis, we identified some EB systems using the generated algorithm and visually defined candidate pulsating component(s) in some. In this section, we would like to distinguish δ Scuti-type oscillation in EBs. To accurately determine the pulsational variability, first, the binary variation has to be removed from the light curve. For this purpose, the method used in the study of Kahraman Aliçavuş et al. (2022) could be used. In this method, the frequency of the orbital period and its harmonics are fitted to the data. Therefore, we need to know the orbital periods of all systems before starting this analysis. Hence, we first determined the orbital period of the targets using the PERANSO Light Curve and Period Analysis Software.[3] PERANSO could be used to calculate the minima times with different methods. We utilized the Kwee Van Woerden method (Kwee & van Woerden 1956) to read the minima times from two consecutive primary eclipses and then calculated the orbital periods for each system. The list of calculated orbital periods ($P_{orb}$) is given in Table 1.

Before removing binarity from the original TESS light curve, we also carried out a contamination analysis. Because of the size of the TESS pixel (21 arcsec) that may consist of the signal from another star(s) placed close to our target. In that case, we could see both eclipse and a pulsation, which could be caused by different objects. Hence, the target pixel files were studied to be sure that all fluxes we take into account come from the target of interest. As we found no significant effect from another source in the pixel files, we continued our analysis. One example of this research is given in Fig. 2.

After the determination of the $P_{orb}$ and being certain of the absence of contamination, the binary variations were removed from the light curve of each system. For this and further analysis, the PERIOD04 software (Lenz & Breger 2005) was used. We carried out a frequency analysis on the residual light curves of each system. Since we want to determine the δ Scuti-type variations, we searched for the frequencies between ∼4–80 d$^{-1}$. The frequencies having signal-to-noise ratio (SNR) over 4.5 were considered as significant (Baran & Koen 2021). Consequently, we found some systems exhibiting δ Scuti-type variations in EBs. However, Chen et al. (2022) published a study and identified around half of our systems (∼60) as EBs with δ Scuti components like us. This study shows our success in classification on δ Scuti stars in eclipsing binaries. Since some systems were defined as δ Scuti stars in EBs in the study of Chen et al. (2022), we excluded these systems from our list and we only list, in our paper, the systems not present in their study. The list of the systems is given in Table 1. In this table, we listed 42 δ Scuti candidates in EBs with their $P_{orb}$ determined in this study and some basic properties taken from the TIC (Stassun et al. 2019). The light curve of an identified system is given in Fig. 3. The calculated frequencies of the highest amplitudes for each target are given in Table 2. The frequency spectrum for one target is also given in Fig. 4. The frequency spectra and light curves of all systems are given in Fig. A1.

## 5 CALCULATION OF BASIC PARAMETERS

We calculated some basic parameters of the determined EBs with δ Scuti component(s), such as absolute magnitude ($M_V$), bolometric magnitude ($M_{bol}$), and luminosity ($L$). The interstellar reddening, $E(B − V)$, was estimated in advance of the calculation of these parameters.

---

[2]This research was made in August 2022

[3]https://www.cbabelgium.com/peranso/

**Table 1.** Information about the identified eclipsing binary systems. $P_{orb}$ values were measured in this study. $T_{eff}$ and log $g$ parameters were taken from TIC (Stassun et al. 2019). * shows the stars that were identified as eclipsing binary systems for the first time in our study.

| ID | TIC number | Other name | RA$^o$ (J2000) | DEC$^o$ (J2000) | Spectral Type | $P_{orb}$ (day) | V (mag ± 0.01) | $T_{eff}$ (K ± 168) | log $g$ ± 0.09 |
|---|---|---|---|---|---|---|---|---|---|
| 1 | 4492575* | HD 218358 | 346.851 | −11.807 | A8IV$^a$ | 39.1817(1) | 7.67 | 7317 | 3.79 |
| 2 | 10756751 | GP Cet | 9.230 | −5.874 | F0V$^b$ | 3.4885(1) | 9.89 | 6703 | 3.46 |
| 3 | 19118532* | TYC 5434-780-1 | 123.426 | −11.558 | – | 4.5868(5) | 11.67 | 6837 | 3.75 |
| 4 | 35481236 | TYC 5949-2585-1 | 101.969 | −16.716 | – | 1.8431(4) | 10.21 | 7803 | 4.03 |
| 5 | 35913152* | UCAC4 226-044751 | 149.323 | −44.969 | – | 5.4778(4) | 12.42 | 7250 | 4.12 |
| 6 | 47175973* | HD 83911 | 145.345 | −6.617 | F2/3V$^b$ | 9.7728(3) | 10.09 | 5887 | – |
| 7 | 66493199 | TYC 6999-424-1 | 15.392 | −30.162 | – | 0.5989(1) | 11.70 | 5323 | 4.30 |
| 8 | 70555928 | AU For | 33.760 | −33.851 | – | 6.1164(3) | 11.21 | 6729 | 3.39 |
| 9 | 89975166 | V707 CrA | 276.125 | −44.199 | F0VkA3mA3$^c$ | 42.5203 | 8.11 | 7546 | 3.65 |
| 10 | 101654574 | HD 191306 | 302.848 | −48.379 | A7V$^d$ | 18.4611(1) | 9.02 | 7500 | 3.89 |
| 11 | 120959196* | TYC 5877-995-1 | 54.319 | −19.641 | – | 0.7852(2) | 12.01 | 7964 | 4.35 |
| 12 | 139699256 | CH Ind | 322.427 | −50.342 | A9V$^d$ | 2.9575(1) | 7.52 | 7023 | 3.43 |
| 13 | 144085463* | V596 Pup | 126.889 | −20.844 | A1V$^a$ | 4.5962(3) | 11.85 | 9311 | – |
| 14 | 152377056 | HD 137752 | 232.359 | −44.544 | A2/3III/IV$^d$ | 0.9433(2) | 8.85 | 7899 | – |
| 15 | 152513129* | TYC 7743-154-1 | 172.107 | −39.730 | – | 2.0634(2) | 12.02 | 7575 | 4.04 |
| 16 | 152796589* | V4437 Sgr | 304.992 | −32.617 | A9IV/V$^e$ | 1.1366(1) | 7.24 | 7663 | 3.84 |
| 17 | 152926076* | TYC 7997-645-1 | 338.422 | −38.890 | – | 0.8033(1) | 10.78 | 6496 | 3.77 |
| 18 | 158536052 | CX Phe | 19.879 | −48.295 | F0V$^d$ | 19.9757(1) | 8.79 | 7048 | 3.87 |
| 19 | 158582033 | UCAC4 195-001257 | 21.454 | −51.170 | – | 0.6512(3) | 12.06 | 7791 | 4.26 |
| 20 | 160708862* | HD 134004 | 227.27 | −42.705 | A1V$^d$ | 3.3449(1) | 9.95 | 9364 | 4.32 |
| 21 | 165456443 | HD 104186 | 179.964 | −36.148 | A5/7m$^e$ | 8.6290(7) | 9.49 | 7372 | 4.26 |
| 22 | 173268508* | TYC 7844-2068-1 | 234.665 | −41.532 | – | 4.7962(2) | 11.09 | 8672 | 4.35 |
| 23 | 175394569* | V634 Vir | 198.906 | −1.208 | – | 0.8682(1) | 11.76 | 7607 | 4.30 |
| 24 | 191926695 | HD 78287 | 136.525 | −38.886 | A3IV/V$^e$ | 1.9349(2) | 9.47 | – | – |
| 25 | 207080350 | HD 203493 | 320.919 | −39.779 | F3IV/V$^d$ | 9.0597(4) | 7.46 | 6978 | 3.98 |
| 26 | 248959865* | HD 8943 | 22.005 | −2.035 | kA9hA9mF1III$^f$ | 44.3267(3) | 7.04 | 7153 | 3.97 |
| 27 | 260407924* | TYC 8183-2363-1 | 152.520 | −45.092 | – | 11.5006(5) | 12.18 | 7062 | 3.75 |
| 28 | 270622446 | AL Scl | 358.819 | −31.921 | B5/8$^e$ | 2.4451(1) | 6.09 | 13211 | – |
| 29 | 279569707* | HD 54011 | 105.968 | −57.629 | A1/2mA5-F0$^g$ | 3.9776(2) | 9.29 | 7801 | – |
| 30 | 311699919* | TYC 9448-1702-1 | 261.617 | −77.832 | – | 2.0378(1) | 10.37 | 7092 | 3.79 |
| 31 | 326751253* | TYC 9013-1359-1 | 210.128 | −65.089 | – | 2.9159(2) | 11.20 | 7186 | 3.51 |
| 32 | 371119381* | HD 134684 | 228.074 | −33.566 | A3/5V$^e$ | 0.7858(1) | 10.00 | 6599 | 3.69 |
| 33 | 375879013* | HD 304971 | 152.408 | −61.411 | A3$^h$ | 0.9197(2) | 9.64 | 7894 | 4.27 |
| 34 | 394517162 | TYC 9106-7-1 | 319.331 | −62.423 | – | 1.6851(3) | 11.81 | 6838 | 4.38 |
| 35 | 402624000* | HD 130267 | 223.412 | −77.327 | A8III$^g$ | 1.7384(3) | 9.21 | 7178 | 3.62 |
| 36 | 409934330 | HD 189995 | 302.148 | −71.552 | A2mA4-F0$^g$ | 2.4890(2) | 7.89 | – | – |
| 37 | 419744996 | HD 205726 | 325.194 | −69.375 | A8V$^g$ | 7.0049(1) | 8.81 | 7553 | 3.54 |
| 38 | 441041386 | HD 206035 | 324.993 | −23.353 | A2$^a$ | 0.7487(1) | 10.64 | 7615 | 4.23 |
| 39 | 448442236* | TYC 9244-354-1 | 210.581 | −68.848 | – | 5.6554(5) | 11.54 | 6454 | 3.71 |
| 40 | 459349787* | TYC 6772-1655-1 | 233.700 | −27.651 | – | 1.5516(1) | 10.29 | 6780 | – |
| 41 | 461080504* | TYC 8957-1789-1 | 159.580 | −61.290 | – | 6.0989(1) | 11.28 | 7317 | 4.14 |
| 42 | 1021067547* | HD 125386A | 215.330 | −61.649 | – | 3.6696(1) | 9.48 | 9669 | – |

*Note.* $^a$Houk & Smith-Moore (1988), $^b$Houk & Swift (1999), $^c$Gray et al. (2017), $^d$Houk (1978), $^e$Houk (1982), $^f$McGahee et al. (2020), $^g$Houk & Cowley (1975), $^h$Nesterov et al. (1995).

To estimate the $E(B − V)$ values we utilized the DUSTMAP python package which consists of two, three-dimensional maps of interstellar dust reddening. First, a three-dimensional dust map of Bayestar19 (Green et al. 2019) was used in the calculations. However, as it has a coordinate limit for the $E(B − V)$ estimation, we could not calculate the $E(B − V)$ values for all of our systems. Therefore, for some stars, we used the Schlegel, Finkbeiner & Davis (1998) far-infrared dust (SFD) map which is a two-dimensional dust map giving the reddening of the whole sky. To convert the SFD of $E(B − V)$, the conversation given in Schlafly & Finkbeiner (2011) was used. The $E(B − V)$ value was then converted to the extinction coefficient ($A_v$) using the same equation given by Poro et al. (2021). The calculated $E(B − V)$ and $A_v$ values are given in Table 3.

Utilizing the $A_v$ values, distance modulation, the bolometric correction taken from Flower (1996) and the Gaia parallaxes (Gaia Collaboration 2021), we calculated the $M_V$, $M_{bol}$, and the $L$ parameters of the systems. In the calculation, the TIC $T_{eff}$ values were taken into account. However, for some targets, there are no TIC $T_{eff}$ values (see Table 1). For them, the spectral types given in Table 1 were considered, and using the calibration given by Gray & Corbally (2009), their $T_{eff}$ values were estimated. The calculated parameters are given in Table 3. These calculated parameters represent the binary systems, not the pulsating stars. However, in this study, we focus on the pulsating component. Therefore, to figure out how much those calculated values represent the binary system or one component we estimated the flux ratios of both binary components using the same method explained by Kahraman Aliçavuş et al. (2022). We measured the eclipse areas that approximately represent the fluxes of binary components. However, one should keep in mind that the measured flux ratios in this study could be different than the real

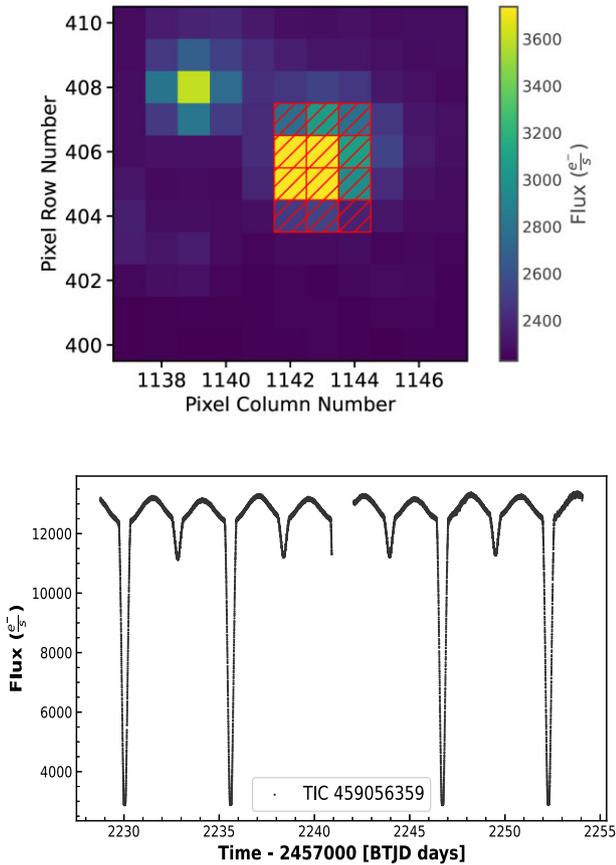

**Figure 2.** Top panel: Pixel files for TIC 459056359. There are two different sources in the frame. The one star covered with hatched red lines is our target. Bottom panel: The flux distribution for only our target star.

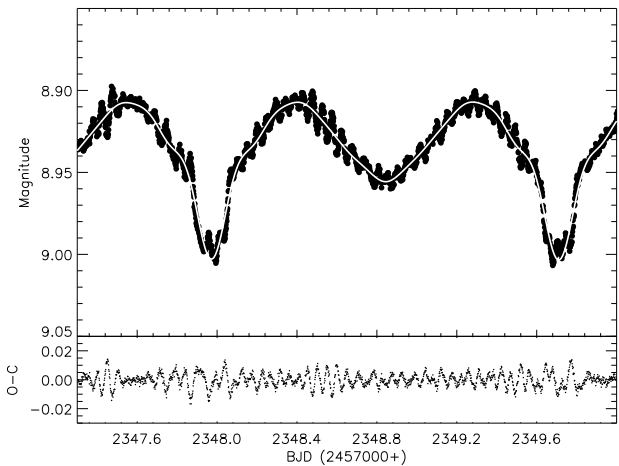

**Figure 3.** Upper panel: the fit (grey line) of the orbital frequency and its harmonics to the TESS data of TIC 402624000. Lower panel: residuals.

light ratios. Especially, if the system has an eccentric orbit and/or the binary components in the system have different radii, the flux ratio would be significantly different than the real light ratio (Torres et al. 2000). In this study, we aimed to estimate rough flux ratios just to show how much the calculated parameters represent the more luminous binary components. The estimated flux ratios are given in

**Table 2.** Frequencies of the highest amplitudes for each target.

| TIC Number | Frequency ($d^{-1}$) | Amplitude (mmag) | Phase |
|---|---|---|---|
| 4492575 | 14.2178 (2) | 0.76 (1) | 0.820 (2) |
| 10756751 | 15.0762 (11) | 0.26 (1) | 0.502 (7) |
| 19118532 | 13.7625 (10) | 0.70 (3) | 0.968 (7) |
| 35481236 | 6.0025 (4) | 0.46 (1) | 0.393 (5) |
| 35913152 | 3.0803 (1) | 0.56 (5) | 0.718 (7) |
| 47175973 | 13.0149 (2) | 3.00 (3) | 0.980 (1) |
| 66493199 | 4.8144 (34) | 0.31 (5) | 0.688 (25) |
| 70555928 | 14.0699 (2) | 2.64 (4) | 0.420 (1) |
| 89975166 | 16.5430 (9) | 2.24 (4) | 0.126 (7) |
| 101654574 | 13.8105 (1) | 2.05 (2) | 0.622 (1) |
| 120959196 | 26.5795 (3) | 2.37 (3) | 0.911 (2) |
| 139825840 | 3.8481 (19) | 0.03 (2) | 0.419 (14) |
| 144085463 | 38.5845 (25) | 0.12 (1) | 0.813 (18) |
| 152377056 | 5.6634 (40) | 0.05 (1) | 0.803 (28) |
| 152513129 | 40.0503 (5) | 1.93 (4) | 0.896 (3) |
| 152796589 | 40.0503 (5) | 1.93 (4) | 0.896 (3) |
| 152926076 | 28.6239 (89) | 0.20 (3) | 0.493 (22) |
| 158536052 | 5.1859 (2) | 1.05 (2) | 0.660 (2) |
| 158582033 | 22.2311 (5) | 2.44 (5) | 0.821 (3) |
| 160708862 | 5.4113 (2) | 0.08 (1) | 0.252 (2) |
| 165456443 | 9.3360 (4) | 1.47 (2) | 0.020 (3) |
| 173268508 | 1.6935 (3) | 1.13 (4) | 0.519 (1) |
| 175394569 | 27.6424 (4) | 1.92 (4) | 0.730 (3) |
| 191926695 | 25.4672 (5) | 0.51 (1) | 0.588 (3) |
| 207080350 | 8.4640 (7) | 0.13 (0) | 0.542 (5) |
| 248959865 | 14.3345 (1) | 2.64 (2) | 0.892 (1) |
| 260407924 | 9.0592 (1) | 10.27 (5) | 0.6020 (9) |
| 270622446 | 42.6193 (7) | 0.16 (0) | 0.330 (5) |
| 279569707 | 12.9196 (21) | 0.12 (9) | 0.354 (12) |
| 311699919 | 20.0309 (5) | 0.65 (2) | 0.784 (4) |
| 326751253 | 7.8363 (2) | 2.01 (2) | 0.137 (1) |
| 371119381 | 12.9521 (4) | 1.04 (2) | 0.620 (3) |
| 375879013 | 32.4246 (5) | 0.45 (1) | 0.240 (4) |
| 394517162 | 8.4530 (40) | 0.25 (2) | 0.861 (27) |
| 402624000 | 20.9683 (1) | 1.01 (1) | 0.003 (2) |
| 409934330 | 17.4381 (4) | 0.41 (1) | 0.723 (2) |
| 419744996 | 15.0052 (1) | 3.85 (2) | 0.509 (1) |
| 441041386 | 37.4067 (7) | 6.36 (2) | 0.486 (5) |
| 448442236 | 4.7797 (35) | 0.28 (5) | 0.730 (27) |
| 459349787 | 11.9171 (3) | 1.10 (1) | 0.779 (2) |
| 461080504 | 5.4561 (51) | 0.29 (4) | 0.377 (36) |
| 1021067547 | 8.1120 (6) | 0.42 (1) | 0.673 (4) |

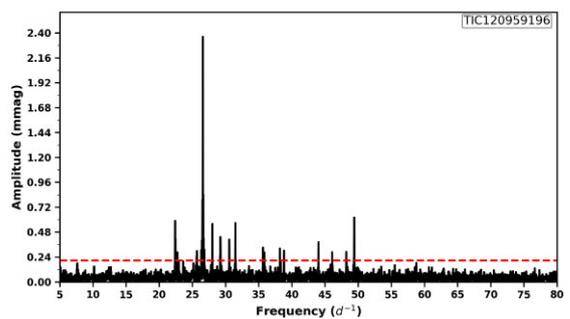

**Figure 4.** The frequency spectrum of TIC 120959196. The dashed horizontal line represents $4.5\sigma$ level.

**Table 3.** Some physical parameters, $E(B-V)$ and $A_v$ values. Superscript $B$ and $S$ represent the used interstellar map of Bayestar2019 and SFD, respectively. The p and s letters stand for primary and secondary. The superscript ∗ presents the flux ratio that cannot be measured or suspected because of pulsation effects on eclipses.

| | TIC Name | $E(B-V)$ (mag ± 0.02) | $A_v$ (mag ± 0.02) | $M_V$ (mag) | $M_{bol}$ (mag) | $\log(L/L_\odot)$ | $I_p/I_s$ ± 0.10 |
|---|---|---|---|---|---|---|---|
| 1 | 4492575 | $0.043^B$ | 0.134 | 1.448 ± 0.032 | 1.483 ± 0.034 | 1.303 ± 0.055 | ∗ |
| 2 | 10756751 | $0.040^B$ | 0.125 | 1.361 ± 0.011 | 1.380 ± 0.067 | 1.344 ± 0.088 | 3.32 |
| 3 | 19118532 | $0.082^B$ | 0.255 | 1.422 ± 0.197 | 1.447 ± 0.199 | 1.317 ± 0.219 | 1.57 |
| 4 | 35481236 | $0.063^B$ | 0.197 | 1.955 ± 0.0063 | 1.986 ± 0.083 | 1.101 ± 0.104 | 0.96 |
| 5 | 35913152 | $0.273^S$ | 0.848 | 2.019 ± 0.070 | 2.054 ± 0.089 | 1.075 ± 0.110 | 1.83 |
| 6 | 47175973 | $0.016^B$ | 0.048 | 1.433 ± 0.067 | 1.378 ± 0.086 | 1.345 ± 0.107 | 1.81 |
| 7 | 66493199 | $0.016^S$ | 0.049 | 4.975 ± 0.213 | 4.851 ± 0.233 | −0.044 ± 0.254 | 3.91 |
| 8 | 70555928 | $0.013^S$ | 0.040 | 1.435 ± 0.088 | 1.455 ± 0.108 | 1.314 ± 0.129 | 2.86 |
| 9 | 89975166 | $0.078^S$ | 0.241 | 0.809 ± 0.033 | 0.839 ± 0.053 | 1.560 ± 0.074 | ∗ |
| 10 | 101654574 | $0.046^S$ | 0.145 | 1.690 ± 0.048 | 1.723 ± 0.068 | 1.207 ± 0.089 | 1.18∗ |
| 11 | 120959196 | $0.055^B$ | 0.171 | 2.492 ± 0.167 | 2.520 ± 0.187 | 0.888 ± 0.208 | 1.04 |
| 12 | 139699256 | $0.020^S$ | 0.063 | 0.921 ± 0.032 | 0.953 ± 0.052 | 1.515 ± 0.073 | 1.07 |
| 13 | 144085463 | $0.017^B$ | 0.054 | −0.298 ± 0.032 | −0.422 ± 0.052 | 2.065 ± 0.074 | 0.87 |
| 14 | 152377056 | $0.266^S$ | 0.824 | −0.406 ± 0.066 | −0.3766 ± 0.096 | 2.047 ± 0.117 | 4.05 |
| 15 | 152513129 | $0.103^S$ | 0.321 | 1.707 ± 0.238 | 1.742 ± 0.268 | 1.199 ± 0.289 | 1.00 |
| 16 | 152796589 | $0.101^S$ | 0.316 | 1.215 ± 0.032 | 1.248 ± 0.062 | 1.397 ± 0.083 | 1.33 |
| 17 | 152926076 | $0.017^S$ | 0.052 | 2.210 ± 0.084 | 2.216 ± 0.086 | 1.010 ± 0.107 | 1.95 |
| 18 | 158536052 | $0.009^S$ | 0.029 | 1.974 ± 0.032 | 2.006 ± 0.052 | 1.093 ± 0.073 | 0.90 |
| 19 | 158582033 | $0.011^S$ | 0.035 | 2.350 ± 0.040 | 2.382 ± 0.060 | 0.943 ± 0.081 | 1.19 |
| 20 | 160708862 | $0.108^S$ | 0.338 | 1.535 ± 0.054 | 1.410 ± 0.074 | 1.332 ± 0.095 | 0.99 |
| 21 | 165456443 | $0.071^S$ | 0.219 | 2.601 ± 0.042 | 2.636 ± 0.062 | 0.842 ± 0.083 | 1.00 |
| 22 | 173268508 | $0.237^S$ | 0.734 | 1.251 ± 0.058 | 1.233 ± 0.078 | 1.403 ± 0.099 | 3.80 |
| 23 | 175394569 | $0.034^B$ | 0.106 | 2.251 ± 0.207 | 2.284 ± 0.227 | 0.982 ± 0.248 | 3.11 |
| 24 | 191926695 | $0.342^S$ | 1.058 | 0.321 ± 0.061 | 0.287 ± 0.124 | 1.787 ± 0.145 | 1.53 |
| 25 | 207080350 | $0.033^S$ | 0.103 | 2.325 ± 0.031 | 2.355 ± 0.051 | 0.954 ± 0.072 | 0.94 |
| 26 | 248959865 | $0.011^B$ | 0.034 | 2.080 ± 0.031 | 2.113 ± 0.051 | 1.051 ± 0.072 | ∗ |
| 27 | 260407924 | $0.158^S$ | 0.489 | 1.244 ± 0.195 | 1.276 ± 0.215 | 1.386 ± 0.236 | 0.47 |
| 28 | 270622446 | $0.013^S$ | 0.040 | −0.650 ± 0.045 | −1.573 ± 0.244 | 2.525 ± 0.266 | 2.16 |
| 29 | 279569707 | $0.101^S$ | 0.313 | 0.778 ± 0.092 | 0.818 ± 0.173 | 1.575 ± 0.194 | 5.53 |
| 30 | 311699919 | $0.196^S$ | 0.607 | 1.623 ± 0.055 | 1.656 ± 0.075 | 1.233 ± 0.096 | 1.72 |
| 31 | 326751253 | $0.880^S$ | 2.727 | −0.880 ± 0.096 | −0.846 ± 0.116 | 2.234 ± 0.137 | 1.16 |
| 32 | 371119381 | $0.249^S$ | 0.772 | 1.763 ± 0.063 | 1.775 ± 0.083 | 1.86 ± 0.103 | 1.23 |
| 33 | 375879013 | $0.324^S$ | 1.005 | 1.463 ± 0.061 | 1.492 ± 0.081 | 1.299 ± 0.102 | 1.38 |
| 34 | 394517162 | $0.247^S$ | 0.767 | 2.931 ± 0.115 | 2.956 ± 0.135 | 0.714 ± 0.156 | 2.65 |
| 35 | 402624000 | $0.728^S$ | 2.257 | −0.749 ± 0.046 | −0.716 ± 0.066 | 2.182 ± 0.087 | 1.11 |
| 36 | 409934330 | $0.248^S$ | 0.044 | 0.280 ± 0.032 | 0.303 ± 0.101 | 1.781 ± 0.123 | 1.05 |
| 37 | 419744996 | $0.728^S$ | 2.257 | −1.597 ± 0.007 | −1.547 ± 0.066 | 2.458 ± 0.087 | 1.34 |
| 38 | 441041386 | $0.062^B$ | 0.193 | 2.194 ± 0.084 | 2.228 ± 0.104 | 1.005 ± 0.125 | 2.47 |
| 39 | 448442236 | $0.247^S$ | 0.767 | 1.569 ± 0.134 | 1.578 ± 0.154 | 1.264 ± 0.176 | 2.40 |
| 40 | 459349787 | $0.148^B$ | 0.460 | 0.546 ± 0.134 | 0.568 ± 0.154 | 1.669 ± 0.175 | 1.00 |
| 41 | 461080504 | $0.624^S$ | 1.935 | 0.693 ± 0.016 | 0.728 ± 0.067 | 1.605 ± 0.088 | ∗ |
| 42 | 1021067547 | $0.018^S$ | 0.055 | 2.212 ± 0.032 | 2.028 ± 0.034 | 1.085 ± 0.055 | 10.11 |

Table 3. As can be seen from the table there are only five systems that have flux ratios over ∼4. For these systems, we may assume that their physical parameters represent mostly the more luminous star.

## 6 DISCUSSION AND CONCLUSIONS

In this study, we present the results of our research on discovering new δ Scuti stars in detached and semi-detached eclipsing binary systems. As a continuation of our previous work (Kahraman Aliçavuş et al. 2022), we searched the southern TESS field using an algorithm developed for this study. This algorithm takes into account the harmonic frequencies of the orbital period, skewness of the light curve, and the UPSILON program to determine the detached and semidetached EB systems. As a result of this research, some EB systems were determined, and the pulsation structure of them was analysed by a visual inspection. In conclusion, we first found 150 EB systems with candidate pulsator(s). Since we aim to find δ Scuti stars in these EBs, a Fourier analysis was carried out on all discovered EB systems after removing binary variations and controlling possible contamination on TESS data. For this purpose, we first estimated the orbital period, $P_{orb}$, and checked the TESS pixel files of each system to find whether there was additional flux contamination coming from another object. As a result, we found no contamination and estimated the $P_{orb}$ for all systems.

We cleaned the binary variation from the TESS light curve and performed a Fourier analysis on the residual to detect δ Scuti-type variation. Consequently, we identified some systems as δ Scuti candidates in EB systems. However, just after our classification, Chen et al. (2022) published their study, and they classified a group of systems (∼60 stars) as δ Scuti stars in EBs, which are also identified the same in our study. Therefore, we decided to list only our new discoveries. In conclusion, we classified 42 objects as new δ Scuti candidates in EB systems. None of these systems are known as δ Scuti pulsators in EBs. Additionally, 25 of the systems listed in Table 1 are identified as eclipsing binary stars for the first time in this study. The distribution of the determined $P_{orb}$, visual magnitude (V), and TIC $T_{eff}$ parameters of the identified systems are given in Fig. 5. According to this figure, most systems have V magnitude lower than $10^m$. So, these systems are suitable for spectroscopic observations. There are also some systems with long $P_{orb}$, while most systems have $P_{orb}$ lower than 5 days. The average $T_{eff}$ value for our systems studied in this paper is around 7500 K. However, one should keep in mind that the $T_{eff}$ values represent the binary system itself not the pulsating component.

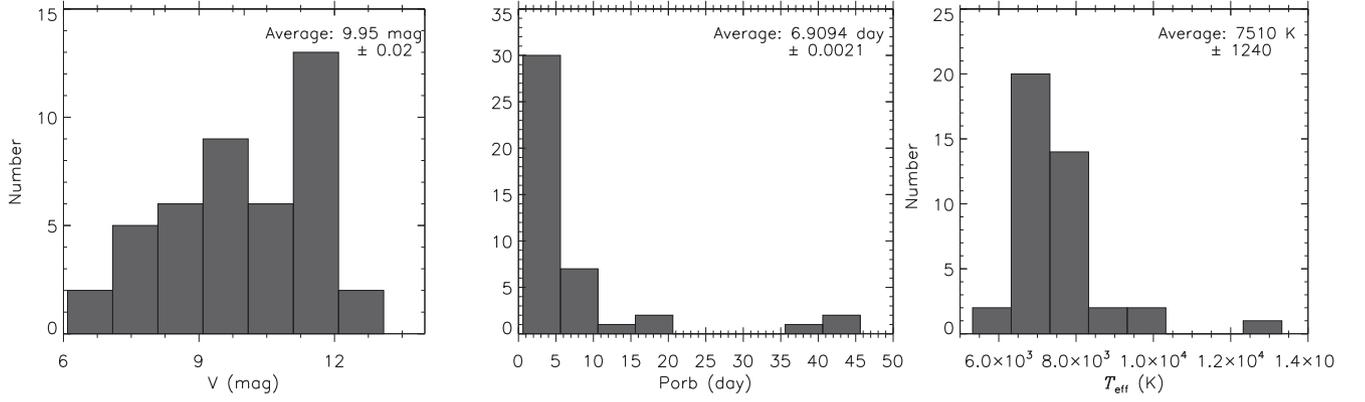

**Figure 5.** The distribution of the determined $P_{orb}$, visual magnitude (V), and TIC $T_{eff}$ parameters.

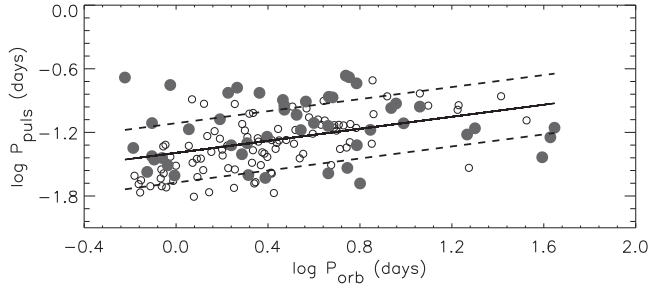

**Figure 6.** The relationship between the orbital ($P_{orb}$) and pulsation ($P_{puls}$) period. Open and filled circles represent known δ Scuti stars in EBs from the study of Kahraman Aliçavuş et al. (2017) and newly discovered ones in this study, respectively. Dashed lines show the standard deviation, while the solid line represents the relationship.

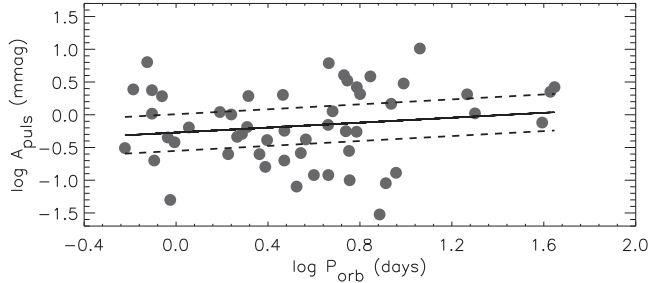

**Figure 7.** The relationship between the $P_{orb} - A_{puls}$. Dashed lines show the standard deviation, while the solid line represents the relationship.

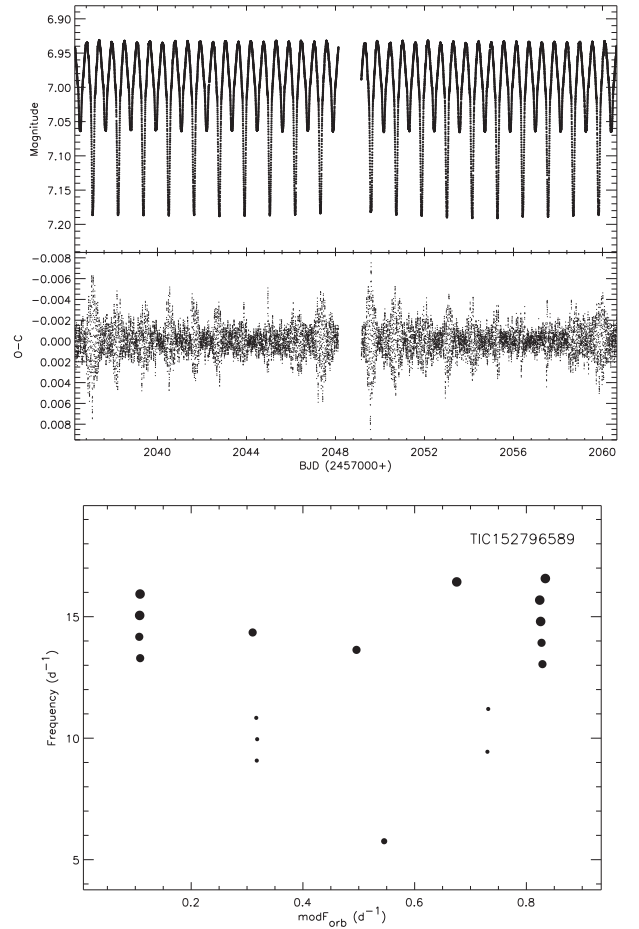

**Figure 8.** Top panel: Short cadence sector 27 data of TIC 152796589 and residuals. The amplitude of pulsations becomes higher in the primary eclipses. Bottom panel: Frequency modulation with the orbital period. The size of the symbols represents the frequency amplitude.

It is known that δ Scuti stars in EBs show different pulsational properties compared to the single δ Scuti stars (Kahraman Aliçavuş et al. 2017). The δ Scuti stars in EBs show some significant relationships such as between the pulsation period ($P_{puls}$) – $P_{orb}$, $P_{orb}$ – Pulsation amplitude ($A_{puls}$), $P_{puls}$ – $T_{eff}$, and $P_{puls}$ – $\log g$. We examined whether our systems obey the $P_{\rm puls} - P_{\rm orb}$ relationship. In Fig. 6, the $P_{\rm puls} - P_{\rm orb}$ relationship is shown for the known $\delta$ Scuti in EB systems and the one discovered in this study. As can be seen from the figure, the systems discovered in this study fit the relationship very well within the standard deviation. The $P_{\rm orb} - A_{\rm puls}$ relationship was checked as well for our stars as given in Fig. 7. In the figure, only our systems were used because our stars and known $\delta$ Scuti stars in eclipsing binaries have $A_{\rm puls}$ values in different filters. As can be seen from the figure, there is no significant relationship between the given parameters. Since we do not have the pulsating components' atmospheric parameters, we cannot investigate other relationships seen in $\delta$ Scuti stars in EB systems.

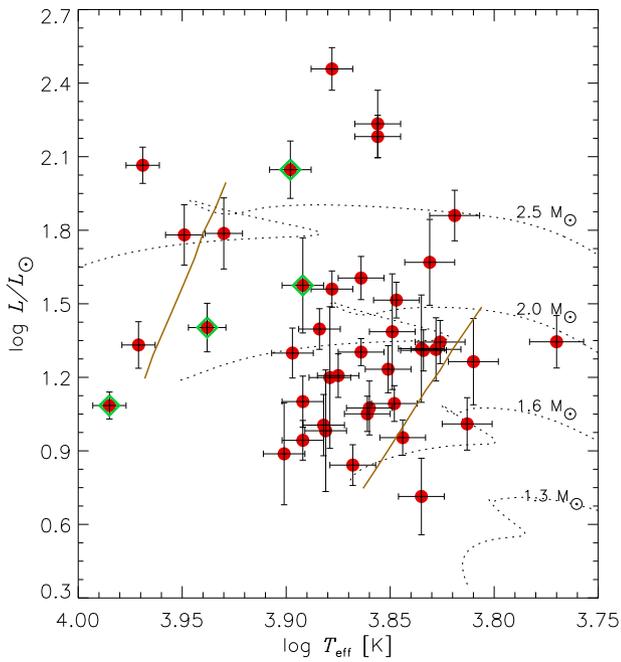

**Figure 9.** Positions of the targets on the H–R diagram. Green dots represent the systems having $I$ ratio higher than $\sim 4$, while red smaller dots show other systems. The solid and dashed lines are the observational borders of the $\delta$ Scuti instability strip (Murphy et al. 2019) and the evolutionary tracks taken from the MESA Isochrones and Stellar Tracks (MIST) (Paxton et al. 2011; Choi et al. 2016; Dotter 2016).

In addition, we made a preliminary investigation to find if there are frequency modulations caused by orbital motion in our systems. We found some systems that show possible frequency modulations; however, since we analysed only one sector of data, the result is insufficient to make a definitive conclusion. Nevertheless, we identified one system as a candidate tidally tilted pulsator, TIC 152796589. The system exhibits both frequency modulation and amplitude modulation by the orbital phase. In the top panel of Fig. 8, it can be observed that the amplitude of the frequencies becomes higher in the primary eclipses compared to the secondary eclipses. Additionally, the frequencies seem to be modulated with the orbital period. Therefore, we classified this system as a candidate tidally tilted pulsator. To achieve certainty about this variability, more TESS data is needed, and amplitude variation through the orbital phases should also be examined (see Fig. 3 in Handler et al. 2020).

The position of the systems is shown on the H–R diagram in Fig. 9. As can be seen from the figure, most systems are placed inside the $\delta$ Scuti instability strip, with most of them clustered near the cool border of the strip. We should keep in mind that the parameters used to build the H–R diagram do not only represent the pulsating components. Therefore, most pulsating components of the systems in this diagram should have different parameters. Depending on the other non-pulsating (or pulsating) component's $T_{\rm eff}$ value, the binary systems could be observed to be cooler or hotter than it appears. Nevertheless, Fig. 9 shows the rough positions of the examined systems.

The pulsating stars in EB systems are critical objects for examining the structure and evolution of the stellar systems. Therefore, it is very essential to increase the sample size of such systems. With this study, we introduce 42 new $\delta$ Scuti candidates in EB systems, and some of them are suitable for further spectroscopic and photometric analysis in order to reveal their accurate properties.

## ACKNOWLEDGEMENTS

We thank Dr. Southworth for his helpful comments that have improved the manuscript. The authors thank Mr. A. R. Awan for his contribution to the paper. This study has been supported by the Scientific and Technological Research Council (TUBITAK) project 120F330. The TESS data presented in this paper were obtained from the Mikulski Archive for Space Telescopes (MAST). The NASA Explorer Program provides funding for the TESS mission. This research made use of Lightkurve, a PYTHON package for Kepler and TESS data analysis (Lightkurve Collaboration 2018). This work has made use of data from the European Space Agency (ESA) mission *Gaia* (http://www.cosmos.esa.int/gaia), processed by the Gaia Data Processing and Analysis Consortium (DPAC; http://www.cosmos.esa.int/web/gaia/dpac/consortium). Funding for the DPAC has been provided by national institutions, in particular, the institutions participating in the Gaia Multilateral Agreement. This research has made use of the SIMBAD data base, operated at CDS, Strasbourg, France.

## DATA AVAILABILITY

The data underlying this work will be shared at reasonable request to the corresponding author.

## APPENDIX A:

**Table A1.** The list of test stars. The abbreviation of ROT represents the rotational variables.

| | TIC number | Type | Vmag (mag) ± 0.01 | Orbital /frequency period (d) | Ref |
|---|---|---|---|---|---|
| 1 | 7724421 | EB | 9.94 | 0.944 | 1 |
| 2 | 9036901 | EB | 12.09 | 3.413 | 1 |
| 3 | 20299889 | EB | 10.19 | 7.606 | 1 |
| 4 | 26801525 | EB | 11.23 | 3.897 | 1 |
| 5 | 41694016 | EB | 10.99 | 0.847 | 1 |
| 6 | 53085065 | EB | 5.20 | 2.933 | 1 |
| 7 | 69819180 | EB | 8.18 | 2.073 | 1 |
| 8 | 84342725 | EB | 6.93 | 2.699 | 1 |
| 9 | 125236734 | EB | 9.32 | 20.178 | 1 |
| 10 | 180570279 | EB | 7.93 | 4.798 | 1 |
| 11 | 200297691 | EB | 9.79 | 6.979 | 1 |
| 12 | 200440175 | EB | 11.86 | 3.652 | 1 |
| 13 | 256352113 | EB | 7.00 | 0.936 | 1 |
| 14 | 269609504 | EB | 11.96 | 1.859 | 1 |
| 15 | 307679747 | EB | 10.20 | 10.173 | 1 |
| 16 | 323143096 | EB | 7.61 | 8.439 | 1 |
| 17 | 323686506 | EB | 11.47 | 1.799 | 1 |
| 18 | 343127696 | EB | 10.59 | 4.672 | 1 |
| 19 | 371480725 | EB | 10.41 | 1.472 | 1 |
| 20 | 399832376 | EB | 10.92 | 1.675 | 1 |
| 21 | 406798603 | EB | 10.26 | 0.906 | 1 |
| 22 | 417175621 | EB | 7.13 | 3.333 | 1 |
| 23 | 428942240 | EB | 8.36 | 1.052 | 1 |
| 24 | 434625997 | EB | 8.15 | 2.058 | 1 |
| 25 | 459056359 | EB | 10.53 | 5.561 | 1 |
| 26 | 1129237 | RR Lyr | 11.53 | 0.511 | 2 |
| 27 | 25894218 | RR Lyr | 13.25 | 0.455 | 2 |
| 28 | 176281431 | RR Lyr | 12.73 | 0.741 | 2 |
| 29 | 246938054 | RR Lyr | 12.77 | 0.567 | 2 |
| 30 | 306392378 | RR Lyr | 11.90 | 0.586 | 2 |
| 31 | 397589987 | RR Lyr | 12.25 | 0.701 | 2 |
| 32 | 7203984 | *δ* Scuti | 10.41 | 0.219 | 3 |
| 33 | 12473170 | *δ* Scuti | 9.05 | 0.056 | 3 |
| 34 | 191466237 | *δ* Scuti | 9.35 | 0.125 | 3 |
| 35 | 327585336 | *δ* Scuti | 7.83 | 0.300 | 3 |
| 36 | 395320456 | *δ* Scuti | 7.24 | 0.170 | 3 |
| 37 | 405251305 | *δ* Scuti | 7.21 | 0.165 | 3 |
| 38 | 435860104 | *δ* Scuti | 6.10 | 0.104 | 3 |
| 39 | 440665786 | *δ* Scuti | 5.21 | 0.069 | 3 |
| 40 | 137099307 | *γ* Doradus | 11.17 | 1.186 | 4 |
| 41 | 219234987 | *γ* Doradus | 4.20 | 0.733 | 5 |
| 42 | 327701133 | *γ* Doradus | 5.00 | 0.217 | 5 |
| 43 | 414641494 | *γ* Doradus | 7.01 | 0.230 | 5 |
| 44 | 408002010 | *γ* Doradus | 7.48 | 0.198 | 5 |
| 45 | 33945685 | ROT | 8.25 | 0.353 | 6 |
| 46 | 38602305 | ROT | 5.87 | 2.976 | 6 |
| 47 | 41331819 | ROT | 5.04 | 1.401 | 6 |
| 48 | 115177591 | ROT | 6.90 | 0.576 | 6 |
| 49 | 139468902 | ROT | 6.92 | 0.455 | 6 |
| 50 | 141281495 | ROT | 8.10 | 2.994 | 6 |

*Notes.* References: 1. Avvakumova, Malkov & Kniazev (2013), 2. Drake et al. (2013), 3. Rodríguez, López-González & López de Coca (2000), 4. Li et al. (2020), 5. Krisciunas & Handler (1995), 6. Balona et al. (2019).

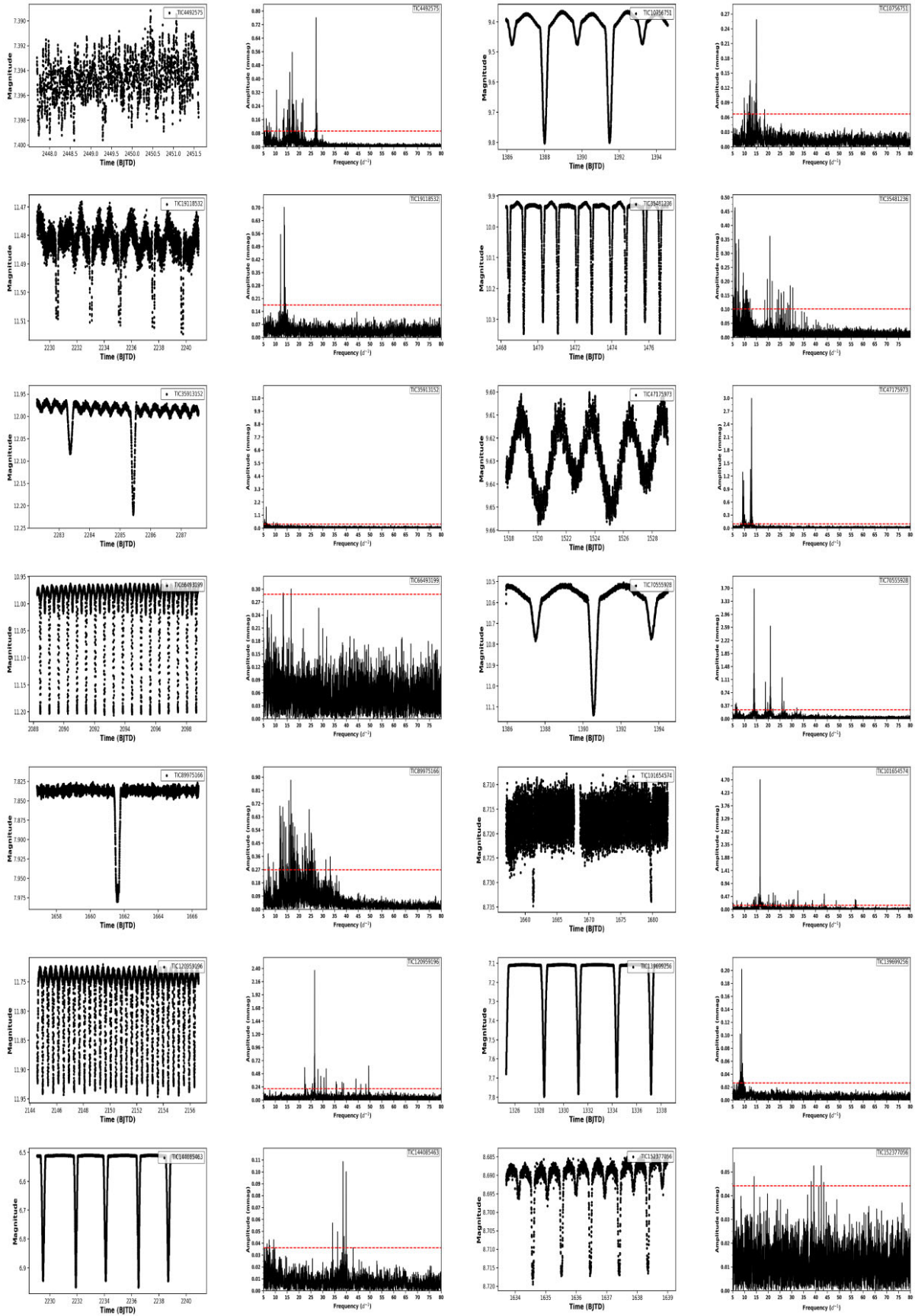

**Figure A1.** TESS light curve and amplitude spectra of the targets.

**Figure A1.** *Continued.*

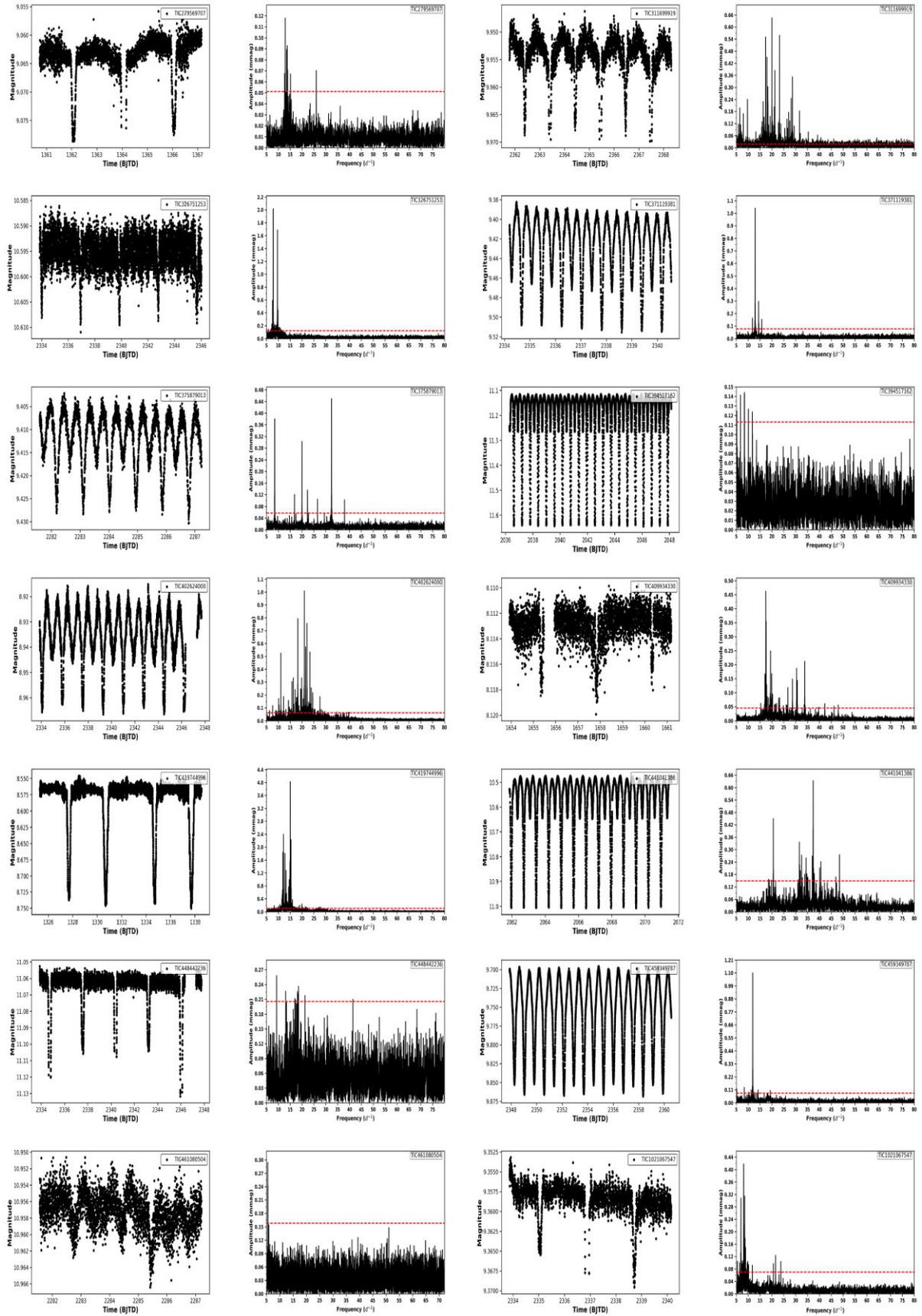

**Figure A1.** *Continued.*